\documentclass[11pt,a4paper]{article}

\usepackage{jheppub,bm}
\usepackage{jheppub}

\usepackage{amsmath,amssymb}
\usepackage{color,hyperref}
\usepackage[T1]{fontenc}
\usepackage{tocloft}
\usepackage[normalem]{ulem}
\usepackage{subcaption}
\usepackage{array}
\usepackage{enumitem}
\usepackage{extarrows}
\usepackage{mathtools}
\usepackage{slashed} 
\usepackage{physics} 

\allowdisplaybreaks[1] 

\newcommand{\eps}{\epsilon}

\newcommand{\ns}{{\slashed{n}}}
\newcommand{\nsb}{{\slashed{\bar{n}}}}
\newcommand{\als}{\alpha_s}

\newcommand{\grid}{{\rm Grid}}
\newcommand{\calC}{{\cal C}}

\newcommand{\df}{{\rm d}}
\newcolumntype{P}[1]{>{\centering\arraybackslash}p{#1}}
\def\calT{\mathcal{T}}

\newcommand{\plus}{\!+\!}
\newcommand{\Li}{{\rm Li}}

\newcommand{\overbar}[1]{\,\overline{\!{#1}}}

\makeatletter
\def\@fpheader{~}
\makeatother

\title{NNLO beam functions for angularity distributions}

\author[a]{Guido Bell,} 
\author[a]{Kevin Brune,} 
\author[b]{Goutam Das,} 
\author[a]{and Marcel Wald~}

\note{SI-HEP-2024-19, TTK-24-34, P3H-24-064.}

\affiliation[a]{
Theoretische Physik 1, Center for Particle Physics Siegen,\\ 
Universit\"at Siegen, 57068 Siegen, Germany}

\affiliation[b]{
Institut f\"ur Theoretische Teilchenphysik und Kosmologie, \\
RWTH Aachen University, 52056 Aachen,
Germany}

\emailAdd{bell@physik.uni-siegen.de}
\emailAdd{brune@physik.uni-siegen.de}
\emailAdd{goutam@physik.rwth-aachen.de}
\emailAdd{marcel.wald@uni-siegen.de}

\abstract{
The popular class of angularity event shapes provides a wealth of information on the hadronic final-state distribution in collider events. While initially proposed for $e^+ e^-$ collisions, angularities have more recently attracted considerable interest as a jet sub\-structure observable at hadron colliders. Moreover, angularities can be measured as a global event shape in deep inelastic electron-nucleon scattering (DIS), and the respective factorisation theorem contains a beam function that parametrises the collinear initial-state radiation. In the present work, we compute the quark and gluon beam functions for seven different angularities to next-to-next-to-leading order (NNLO) in the strong-coupling expansion.  Our calculation is based on an automated framework that was previously developed for SCET-2 observables, and which we transfer in the current work to the generic SCET-1 case. Our results are relevant for resumming DIS angularity distributions at NNLL$'$ accuracy.  
}


\begin{document} 

\maketitle

\flushbottom

%%%%%%%%%%%%%%%%%%%%%%%%%%%%%%%%%%%%%%%%%%%
\section{Introduction}
%%%%%%%%%%%%%%%%%%%%%%%%%%%%%%%%%%%%%%%%%%%
\label{sec:introduction}

Event-shape variables characterise the hadronic final-state distribution in collider events. Due to their distinctive simplicity, they are widely used for precision studies of the QCD dynamics, for extracting the strong-coupling constant or for tuning parton showers and hadronisation models in Monte-Carlo event generators~\cite{Dasgupta:2003iq}. The specific class of angularity distributions, in particular, provides a handle to control the relative importance of different dynamical contributions as a function of a continuous parameter $A<2$~\cite{Berger:2003iw}. Angularities have been extensively studied at $e^+e^-$ colliders (see e.g.~\cite{Hornig:2009vb,Almeida:2014uva,Procura:2018zpn,Banfi:2018mcq,Bell:2018gce,Budhraja:2019mcz}), in Higgs-boson decays~\cite{Zhu:2023oka,Yan:2023xsd} and as a probe of the internal structure of jets at hadron colliders (see e.g.~\cite{Almeida:2008yp,Ellis:2010rwa,Larkoski:2014uqa,Larkoski:2014pca,Hornig:2016ahz,Kang:2018vgn,Caletti:2021oor,Reichelt:2021svh,Dasgupta:2022fim,Chien:2024uax}).

In view of the planned Electron-Ion Collider~\cite{AbdulKhalek:2021gbh,AbdulKhalek:2022hcn,Abir:2023fpo}, event-shape distributions in deep inelastic electron-nucleon scattering (DIS) have  received renewed interest recently~\cite{Kang:2013nha,Kang:2013lga,Kang:2014qba,Gehrmann:2019hwf,Aschenauer:2019uex,Li:2020bub,Zhu:2021xjn,Chu:2022jgs,Knobbe:2023ehi,Cao:2024ota,Fang:2024auf}. Specifically, DIS angularities were introduced as a global (dimensionless) measure in~\cite{Zhu:2021xjn}, and we follow these definitions and conventions here. In terms of two massless reference vectors $q_B^\mu$ and $q_J^\mu$, which specify the beam and a (pre-determined) jet axis, the angularity is defined as 
\begin{align}
\calT_A(\lbrace k_{i} \rbrace) =
 \frac{2}{Q^2} \,\sum_i\, \bigg\{ \theta(q_J\cdot k_i - q_B\cdot k_i)\,
 (q_B\cdot k_i)\, \bigg(\frac{q_B\cdot k_i}{q_J\cdot k_i} \bigg)^{-A/2} + (q_B^\mu \leftrightarrow q_J^\mu) \bigg\}\,,
\end{align}
where $q^2=-Q^2$ is the momentum transfer, and $\lbrace k_{i} \rbrace$ denote the momenta of the final-state particles on which the angularity is measured. For small values $\calT_A\ll 1$, it was furthermore shown in~\cite{Zhu:2021xjn} that the differential angularity distribution satisfies a factorisation theorem,
\begin{align}
\frac{\df \sigma}{\df x \,\df Q^2 \df \calT_A} &= 
\frac{\df \sigma_0}{\df x \,\df Q^2} \;
\sum_{i,j} \,H_{ij}(Q,\mu)
 \int \df \calT_A^J \; \df \calT_A^B\; \df \calT_A^S\;  
\delta\Bigl(\calT_A-\calT_A^J-\calT_A^B-\calT_A^S\Bigr)
\nonumber\\
&\times  B_{i/h}\big(x,\calT_A^B,\mu\big)\; 
J_j\big(\calT_A^J,\mu\big) \; S_{ij}\big(\calT_A^S,\mu\big)\,,
\end{align}
which is valid for angularities with $A<1$ up to power corrections of $\mathcal{O}(\calT_A)$. Here $\sigma_0$ is the Born cross section, $x$ denotes the Bj\"orken variable, and the sum runs over all partonic channels. Whereas the hard ($H_{ij}$), jet ($J_j$), and soft ($S_{ij}$) functions are related to those that appear in $e^+ e^-$ scattering, the angularity beam function $B_{i/h}$ is a new ingredient that describes collinear initial-state radiation of a parton $i$ in a hadron $h$. In contrast to the former, which are currently available at next-to-next-to-leading order (NNLO) in perturbation theory~\cite{Hornig:2009vb,Bell:2018vaa,Bell:2018oqa,Bell:2021dpb,Brune:thesis}, the beam function has so far only been determined for quark-induced processes to NLO accuracy~\cite{Zhu:2021xjn}.\footnote{For the special case of $A=0$, the beam functions have been determined to N$^3$LO~\cite{Stewart:2010qs,Berger:2010xi,Gaunt:2014xga,Gaunt:2014cfa,Ebert:2020unb,Baranowski:2022vcn}.} The goal of the present article consists in extending this calculation for both the quark and the gluon beam function to NNLO as well. In particular, this allows for a precision resummation of DIS angularity distributions at the primed next-to-next-to-leading logarithmic (NNLL$^\prime$) accuracy.

Our calculation is based on an automated framework that was initially developed for the calculation of NNLO soft functions~\cite{Bell:2018vaa,Bell:2018oqa,Bell:2020yzz,Bell:2023yso}, and later extended to NNLO jet and beam functions~\cite{Bell:2021dpb,Bell:2022nrj,Bell:2022tmi,Brune:thesis,Wald:thesis,Bell:2024epn}. Specifically, we showed in~\cite{Bell:2024epn} that the perturbative beam-function matching kernels can be directly sampled in momentum space, without the need to perform an additional Mellin transform. Whereas the example we considered in~\cite{Bell:2024epn} belongs to the class of SCET-2 observables, we will show in this work that the formalism carries over to generic SCET-1 observables, although the distribution structure is slightly more complicated in this case. Apart from its phenomenological relevance, our calculation therefore also serves another purpose, as it demonstrates that our automated framework for the calculation of NNLO beam functions is now complete.

The remainder of the paper is organised as follows. We will first review the formal definition of the beam functions, as well as their perturbative components and their renormalisation properties in Sec.~\ref{sec:framework}. We will then discuss some technical aspects of the calculation in Sec.~\ref{sec:computation}, before we present our numerical results for seven different angularities in Sec.~\ref{sec:results}. We finally conclude in Sec.~\ref{sec:conclusion}, and collect some details of our study in the appendix.

%%%%%%%%%%%%%%%%%%%%%%%%%%%%%%%%%%%%%%%%%%%
\section{Theoretical framework}
%%%%%%%%%%%%%%%%%%%%%%%%%%%%%%%%%%%%%%%%%%%
\label{sec:framework}

The beam functions are defined as hadronic matrix elements of collinear field operators, $\chi(x) = W^{\dagger}_{\bar n}(x)\frac{\ns \nsb}{4} \psi_c(x)$ and $\mathcal{A}_{c,\perp}^{\mu}(x)=1/g_s \, W_{\bar n}^{\dagger}(x)\big[i D^{\mu}_{c,\perp} \,W_{\bar n} (x)\big]$, where $W_{\bar n}$ denotes a collinear Wilson line, and we use the standard conventions for light-cone coordinates with $n^2=\bar n^2=0$, $n\cdot \bar n =2$, while any transverse four-vector satisfies $k_{\perp}\cdot n =k_{\perp}\cdot \bar n=0$. We furthermore adopt the notation $k^- = \bar n \cdot k$ and $k^+ = n \cdot k$ in this work. For an hadronic state $\ket{h(P)}$ with collinear momentum $P^\mu=P^- n^\mu/2$, the quark beam function is then defined by
\begin{align}
\label{eq:definition:quark}
    \frac12  \left[\frac{\ns}{2}\right]_{\beta \alpha}
    B_{q/h}(x,\tau,\mu) =& \sum_{X} \,
    \delta\Big( (1-x) P^- - \sum_i k_i^- \Big)\,
		{\cal M}(\tau;\{k_i\}) 
		\nonumber\\
    & \times \bra{h(P)}\overbar{\chi}_{\alpha}(0) \ket{X} 
    \bra{X}\chi_{\beta}(0)  \ket{h(P)} , 
\end{align}
where the sum $X$ includes an integration over the phase space of the collinear partons with momenta $\{k_i\}$. Likewise, the gluon beam function is defined by
\begin{align}
\label{eq:definition:gluon}
    B_{g/h}(x,\tau,\mu) =&  -(x P^-) \sum_{X} \,
    \delta\Big( (1-x) P^- - \sum_i k_i^- \Big)\,
		{\cal M}(\tau;\{k_i\}) 
		\nonumber\\
    & \times \bra{h(P)}\mathcal{A}_{c,\perp}^{\mu}(0) \ket{X} 
    \bra{X} \mathcal{A}_{c,\perp,\mu}(0)  \ket{h(P)} . 
\end{align}
These definitions are valid for arbitrary observables, and we will specify the measurement function ${\cal M}(\tau;\{k_i\})$ for DIS angularities in the following section. In general, we assume that the observable-specific distributions are resolved by a Laplace transform, and the \mbox{argument $\tau$} of the beam function refers to the Laplace-conjugate variable of the observable. It is furthermore supposed to have the dimension $1/$mass.

The beam functions are not directly accessible in perturbation theory, but as long as the intrinsic scale of the collinear radiation is perturbative, i.e.~$\tau\ll1/\Lambda_{\rm QCD}$, they can be matched onto the standard parton distribution functions $f_{i/h}(x,\mu)$ via
\begin{align}
\label{eq:matching}
B_{i/h}(x,\tau,\mu) &=\sum_k 
    \int_x^1 \frac{\df z}{z} \,
    I_{i\leftarrow k}(z,\tau,\mu) 
    ~ f_{k/h}\Big(\frac{x}{z},\mu\Big)\,, 
\end{align}
which holds at leading power in $\tau\Lambda_{\rm QCD}\ll1$. In practice, the matching is most efficiently performed using partonic on-shell states, since the parton distribution functions then evaluate to $f_{i/j}(x,\mu)=\delta_{ij}\delta(1-x)$ to all orders in perturbation theory when dimensional regularisation is used for both ultraviolet (UV) and infrared (IR) divergences. The partonic calculation therefore directly yields the matching kernels in this case.

Following~\cite{Bell:2024epn}, the bare matching kernels can then be renormalised using two different counter\-terms that either subtract the UV divergences of the beam function ($Z_i^B$) or the IR divergences associated with the matching onto the parton distribution functions ($Z_{k\leftarrow j}^f$),
\begin{align}
    {I}_{i\leftarrow j}(z,\tau,\mu) = Z_i^B(\tau,\mu)\, 
		\sum_k\int_z^1 \frac{\df z'}{z'} \,
    I_{i\leftarrow k}^{\rm bare}(z',\tau) 
    ~ Z_{k\leftarrow j}^f\Big(\frac{z}{z'},\mu\Big)\,. 
\end{align}
The (observable-independent) IR counterterm satisfies the renormalisation group (RG) equation
\begin{align}
  \frac{\df}{\df\ln \mu} \;
	Z_{k\leftarrow j}^f(z,\mu)
  =&
  -2 \sum_{l}  \int_z^1 \frac{\df z'}{z'} \,
    Z_{k\leftarrow l}^f(z',\mu) 
    ~ P_{l\leftarrow j}\Big(\frac{z}{z'},\alpha_s\Big)\,, 
\end{align}
where $P_{i\leftarrow j}(z,\als)=\sum_m \left(\frac{\alpha_{s}}{4 \pi} \right)^{m+1} P_{i\leftarrow j}^{(m)}(z)$ are the DGLAP splitting functions, whose explicit expressions up to the considered two-loop order were collected in App.~A of~\cite{Bell:2024epn}. The solution of the RG equation is then given by
\begin{align}
&Z_{k\leftarrow j}^f(z,\mu)  =  
\delta_{kj}\,\delta(1-z) + \left( \frac{\alpha_s}{4 \pi} \right) 
\left\{ P_{k\leftarrow j}^{(0)}(z)\,\frac{1}{\eps}
\right\}
\nonumber\\[0.2em]  
&\quad
+\left( \frac{\alpha_s}{4 \pi} \right)^2 
\bigg\{ 
- P_{k\leftarrow j}^{(0)}(z)\,
\frac{\beta_0}{2\eps^2}
+\frac{1}{2\eps^2} \Big( P_{k\leftarrow l}^{(0)} \otimes
 P_{l\leftarrow j}^{(0)} \Big) (z) 
+ P_{k\leftarrow j}^{(1)}(z)\,\frac{1}{2\eps}\bigg\}
\,, 
\label{eq:Zfren}
\end{align}
where $\eps=(4-d)/2$ is the dimensional regulator and $\beta_0 = \frac{11}{3} C_A - \frac43T_F n_f$ is the one-loop coefficient of the QCD $\beta$-function. We also introduced a short-hand notation for the convolutions
\begin{align}
    \Big( P_{k\leftarrow l}^{(0)} \otimes
 P_{l\leftarrow j}^{(0)} \Big) (z) &=
 \sum_{l}  \int_z^1 \frac{\df z'}{z'} \,
    P_{k\leftarrow l}^{(0)}(z') 
    ~ P_{l\leftarrow j}^{(0)}\Big(\frac{z}{z'}\Big) 
\label{eq:def:convolution}    
\end{align}
that appear frequently in the calculation. Explicit expressions for these convolutions can also be found in App.~A of~\cite{Bell:2024epn}.

The RG equation of the UV counterterm, on the other hand, depends on the observable through a parameter $n$ that controls the sensitivity of the observable in the soft-collinear region. For the considered angularity distributions with $A<1$, the observable belongs to the SCET-1 class, and the parameter $n=1-A$ is non-zero. Specifically, the RG equation takes the form
 \begin{align}
  \frac{\df}{\df\ln \mu} \;
	Z_i^B(\tau,\mu)
  =&
  \left[
    2g(n)\Gamma_{\rm cusp}^{R_i}(\als) \, L + \gamma^B_i(\als)
  \right] Z_i^B(\tau,\mu) \,, 
\end{align}
where $g(n) = (n+1)/n$, $L=\ln \big( \mu\bar{\tau}/\left(q_-\bar{\tau}\right)^{1/g(n)}\big)$, $\bar\tau=\tau e^{\gamma_E}$ and we have traded the large component $P_-$ in the beam-function definition by the partonic component $q_-=z P_-$ that enters the hard interaction. The RG equation is controlled by the cusp anomalous dimension $\Gamma_{\rm cusp}^{R_i}(\als)$ in the representation of the parton $i$, and the (observable-dependent) non-cusp anomalous dimension $\gamma^{B}_i(\als)$, which we expand as
\begin{align}
\label{eq:AD} 
\Gamma_{\mathrm{cusp}}^{R_i}(\alpha_s) = \sum_{m=0}^{\infty} \left(\frac{\alpha_{s}}{4 \pi} \right)^{m+1} \Gamma_{m}^i\,, \qquad 
\gamma^{B}_i(\alpha_s)= \sum_{m=0}^{\infty} \left(\frac{\alpha_{s}}{4 \pi} \right)^{m+1} \gamma^{B}_{i,m}\,.
\end{align} 
The two-loop solution of the RG equation then becomes
\begin{align}
&Z_i^B(\tau,\mu)  =  
1 + \left( \frac{\alpha_s}{4 \pi} \right) 
\left\{ - g(n) \frac{\Gamma_0^i}{2\eps^2} - 
\bigg( g(n) \Gamma_0^i L +\frac{\gamma_{i,0}^B}{2} \bigg) \frac{1}{\eps} 
\right\} +\left( \frac{\alpha_s}{4 \pi} \right)^2 
\bigg\{ g(n)^2 \frac{(\Gamma_0^i)^2}{8\eps^4}
\nonumber\\[0.2em]  
 &\quad
+ \left( g(n) \frac{\Gamma_0^i}{2}L
+   \frac{\gamma_{i,0}^B}{4}
 + \frac{3\beta_0}{8}  \right) \frac{g(n)\Gamma_0^i}{\eps^3}
+\bigg( g(n)^2 \frac{(\Gamma_0^i)^2}{2}L^2 
+ g(n) \frac{\Gamma_0^i}{2} \Big(
 \gamma_{i,0}^B + \beta_0 \Big)L
\nonumber\\[0.2em]  
 &\qquad\;
-\frac{g(n) \Gamma_1^i}{8} 
+ \frac{(\gamma_{i,0}^B)^2}{8}
+ \frac{\beta_0\gamma_{i,0}^B}{4}
\bigg) \, \frac{1}{\eps^2}
 - \bigg( g(n) \frac{\Gamma_1^i}{2} L + \frac{\gamma_{i,1}^B}{4} \bigg) \frac{1}{\eps}
\bigg\}
\,, 
\end{align}
and the specific values for the coefficients of the anomalous dimensions $\Gamma_m^i$ and $\gamma_{i,m}^B$ are summarised explicitly for the angularity distributions in App.~\ref{app:anoD}. 

With the RG equations of the counterterms at hand, we next address the renormalised matching kernels ${I}_{i\leftarrow j}(z,\tau,\mu)$, whose scale dependence is controlled by the RG equation
\begin{align}
\label{eq:rgeI}
  \frac{\df}{\df\ln \mu} \;I_{i\leftarrow j}(z,\tau,\mu)
  &=
  \left[
    2g(n)\Gamma_{\rm cusp}^{R_i}(\als) \, L + \gamma^{B}_i(\als)
  \right] I_{i\leftarrow j}(z,\tau,\mu)
	\nonumber\\
  &\quad -2 \sum_{k} \int_z^1 \frac{\df z'}{z'} \,
    I_{i\leftarrow k}(z',\tau,\mu) 
    ~ P_{k\leftarrow j}\Big(\frac{z}{z'},\alpha_s\Big)\,,
\end{align}
which is solved by
\begin{align}
\label{eq:Iren}
&I_{i\leftarrow j}(z,\tau,\mu)  
\\[0.2em]  
&= 
\delta_{ij}\,\delta(1-z) + \left( \frac{\alpha_s}{4 \pi} \right) 
\bigg\{ \Big( g(n)\Gamma_0^i \,L^2 
+\gamma_{i,0}^B \,L \Big) \delta_{ij}\,\delta(1-z)  
- 2L\,P_{i\leftarrow j}^{(0)}(z)
+ I_{i\leftarrow j}^{(1)}(z) \bigg\}
\nonumber\\[0.2em]  
  &
+\left( \frac{\alpha_s}{4 \pi} \right)^2 
\bigg\{ \bigg( \frac{g(n)^2(\Gamma_0^i)^2}{2} L^4 
+g(n)\Gamma_0^i\left( \gamma_{i,0}^B + \frac{2\beta_0}{3} \right) 
L^3 
\nonumber\\[0.2em]  
 &\qquad\qquad\qquad\quad
+ \big( g(n) \Gamma_1^i + \frac12(\gamma_{i,0}^B)^2 
 +\beta_0 \gamma_{i,0}^B  \big) L^2  
+ \gamma_{i,1}^B L\bigg) \delta_{ij}\,\delta(1-z)
\nonumber\\[0.2em]  
 &\qquad\qquad\quad
-2\Big( g(n)\Gamma_0^i L^3 + \big( \beta_0 + \gamma_{i,0}^B\big) L^2\Big) P_{i\leftarrow j}^{(0)}(z)
 + \Big( g(n)\Gamma_0^i  L^2 +(\gamma_{i,0}^B +2\beta_0) L \Big)\,
I_{i\leftarrow j}^{(1)}(z)  
\nonumber\\[0.2em]  
 &\qquad\qquad\quad
 +2 L^2 \Big( P_{i\leftarrow k}^{(0)} \otimes P_{k\leftarrow j}^{(0)}  \Big)(z)
-2 L \Big( I_{i\leftarrow k}^{(1)} \otimes  
P_{k\leftarrow j}^{(0)} \Big)(z)  
- 2L\,P_{i\leftarrow j}^{(1)}(z)
+  I_{i\leftarrow j}^{(2)}(z) \bigg\} . 
\nonumber
\end{align}
As the anomalous dimensions are known to the considered two-loop order, we will focus on the non-logarithmic terms $I^{(m)}_{i \leftarrow j}(z)$ in the following, which we sample numerically for various values of the momentum fraction $z$ for seven different angularities $A$. For completeness, we also give the convolutions between the NLO matching kernels $I_{i\leftarrow k}^{(1)}$ and the one-loop splitting functions $P_{k\leftarrow j}^{(0)}$ explicitly in App.~\ref{app:anoD}.

%%%%%%%%%%%%%%%%%%%%%%%%%%%%%%%%%%%%%%%%%%%
\section{Computational details}
%%%%%%%%%%%%%%%%%%%%%%%%%%%%%%%%%%%%%%%%%%%
\label{sec:computation}

Our calculation is based on an automated framework that was developed in~\cite{Bell:2021dpb,Bell:2022nrj,Bell:2022tmi,Wald:thesis,Bell:2024epn}, and we will not repeat the details of this approach here. Specifically, one starts from the Laplace transform of the angularity beam function, which brings the measurement function in the definition of the beam functions \eqref{eq:definition:quark} and \eqref{eq:definition:gluon} for a single collinear emission into the form
\begin{align}
\label{eq:jet:measure:one-emission}
  {\cal M}_1(\tau;k) 
  &= 
  \exp 
  \left[ 
        -\tau k_T 
        \left( 
          \frac{k_T}{(1-z) P^-}
        \right)^n
        f(t_k)
  \right],
\end{align}
where $k_T = |\vec{k}^\perp|$ refers to the transverse momentum of the emitted parton with respect to the beam axis, and $t_k = (1-\cos \theta_k)/2$ parametrises a non-trivial azimuthal dependence of the observable around this axis (which does not exist for the angularities considered here). The light-cone components of the emitted parton are furthermore set by $k^+=k_T^2/k^-$ and $k^- = (1-z) P^-$. From \eqref{eq:jet:measure:one-emission}, it is evident that the Laplace variable $\tau$ has dimension $1/$mass, and the remaining quantities in this expression are given for the angularity distributions by $n=1-A$ and $f(t_k)=1$.

For two emissions with momenta $k^\mu$ and $l^\mu$, we write similarly
\begin{align}
  {\cal M}_2(\tau; k,l) 
  &= 
  \exp 
  \left[ 
        -\tau q_T 
        \left( 
          \frac{q_T}{(1-z) P^-}
        \right)^n
        {\cal F}(a,b,t_k, t_l, t_{kl})
  \right],
\end{align}
where the various kinematic variables refer to
\begin{align}
\label{eq:parametrisation:two}
  a = \frac{k^- l_T}{l^- k_T}, \qquad
  b = \frac{k_T}{l_T}, \qquad
  \bar z = \frac{k^- + l^-}{P^-}, \qquad
  q_T = \sqrt{(k^- + l^-)(k^+ + l^+)}\,, 
\end{align}
with $\bar z=1-z$, and the observable is now in general characterised by three non-trivial angular variables $t_i = (1-\cos \theta_i)/2$ for $i\in\{k,l,kl\}$, see Sec.~3.3. of~\cite{Bell:2018oqa} for further details. In this notation, the two-emission angularity measurement function becomes
\begin{align}
\label{eq:M2}
 {\cal F}(a,b,t_k, t_l, t_{kl})
&=\frac{a + a^A b}{a + b} \;
\bigg(\frac{a + b}{a(1 + a b)} \bigg)^{A/2}\,, 
\end{align}
which is, in particular, independent of the angular variables. The remainder of the calculation then follows as described in our earlier work (see in particular~\cite{Wald:thesis}), with the most difficult part consisting in isolating the implicit divergences of the phase-space integrations. In comparison to our previous work on jet-veto resummation~\cite{Bell:2022nrj,Bell:2024epn}, an additional complication arises here, since it is crucial in our setup that the measurement function \eqref{eq:M2} stays finite and non-zero in all singular limits of the matrix elements. As this is not automatically the case in the parametrisations we are using, we need to rescale some of the integration variables and to perform additional sector-decomposition steps to bring the measurement function into the desired form (see~\cite{Wald:thesis} for further details). The specific rescalings depend on the value of $n$, and they divide the observables into three classes with $n<1$, $n>1$, and $n=1$ (the latter case not requiring any rescalings). 

Whereas the observable-dependent distributions are resolved by the Laplace transform, the beam-function matching kernels are distribution-valued in the momentum fraction $z$ carried by the parton that enters the hard interaction. Specifically, the distributions arise by rewriting
\begin{align}
(1-z)^{-1+m \eps}=  \frac{\delta(1-z)}{m \eps} 
+
\sum_{k=0}^{\infty}\, \frac{(m \eps)^k}{k!} \,\bigg[\frac{\ln^k (1-z)}{1-z}\bigg]_+ 
\end{align}
and the final result consists of these distributions multiplied by numerical coefficients and a non-trivial $z$-dependent `grid' contribution that we sample for different values of $z$. In order to perform the expansion in the dimensional regulator $\eps$ and the subsequent numerical integrations, we have implemented our setup in the publicly available program {\tt pySecDec}~\cite{Borowka:2017idc}, and we use the {\tt Vegas} routine of the {\tt Cuba} library~\cite{Hahn:2004fe} for the numerical integrations.

%%%%%%%%%%%%%%%%%%%%%%%%%%%%%%%%%%%%%%%%%%%
\section{Results}
%%%%%%%%%%%%%%%%%%%%%%%%%%%%%%%%%%%%%%%%%%%
\label{sec:results}

We will now present our results for the non-logarithmic contributions to the renormalised matching kernels $I_{i\leftarrow j}^{(m)}(z)$ defined in \eqref{eq:Iren} at NLO ($m=1$) and NNLO ($m=2$) for both the quark ($i=q$) and gluon ($i=g$) channels. Using the charge conjugation invariance of QCD, one can immediately derive the corresponding anti-quark kernels ($i=\bar q$) from these expressions.

The NLO kernels can be obtained analytically for any value of the angularity $A<1$. Specifically, they read
\begin{align}
	I_{q \leftarrow q}^{(1)}(z) &= 
	C_F \,\bigg\{
	\frac{A(4-A)}{(1-A)(2-A)}\, \frac{\pi^2}{6} \,\delta (1-z) 
	+ \frac{8(1-A)}{(2-A)} \left[\frac{\ln(1-z)}{1-z}\right]_+   
	\nonumber\\
	&\qquad	\quad\;
	- \frac{4(1-A)}{(2-A)} \bigg( (1+z) \,\ln \Big( \frac{1-z}{z} \Big) + \frac{2}{1-z} \,\ln z \bigg) + 2(1-z)
	\bigg\}\,,
	\nonumber\\[0.4em]
	I_{q \leftarrow g}^{(1)}(z) &= 
	T_F \,\bigg\{
	\frac{4(1-A)}{(2-A)} \, \big(1-2z+2z^2\big) \ln \Big( \frac{1-z}{z} \Big) 
	+ 4 z(1-z)
	\bigg\}\,,
	\nonumber\\[0.4em]
	I_{g \leftarrow q}^{(1)}(z) &= 
	C_F \,\bigg\{ 
	\frac{4(1-A)}{(2-A)} \; \frac{1+(1-z)^2}{z} \, \ln \Big( \frac{1-z}{z} \Big) 
	+2 z 
	\bigg\}\,,
	\nonumber\\[0.4em]
	I_{g \leftarrow g}^{(1)}(z) &= 
	C_A \,\bigg\{
	\frac{A(4-A)}{(1-A)(2-A)}\, \frac{\pi^2}{6} \,\delta (1-z) 
	+ \frac{8(1-A)}{(2-A)}  \left[\frac{\ln(1-z)}{1-z}\right]_+   
	\nonumber\\
	&\qquad	\quad\;
	- \frac{8(1-A)}{(2-A)} \bigg( \frac{z^3-(1-z)^2}{z} \,\ln \Big( \frac{1-z}{z} \Big) + \frac{1}{1-z} \,\ln z \bigg) 
	\bigg\}
	\,. 
	\label{eq:I1ij}
\end{align}
The results for the first two kernels are in agreement with a previous determination~\cite{Zhu:2021xjn}, whereas the latter two kernels have not been presented before.

\begin{figure}[t!]
	\centerline{
		\includegraphics[height=5.2cm]{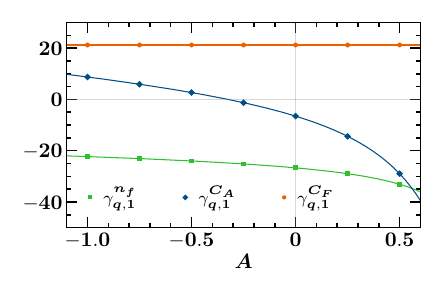}
		\includegraphics[height=5.2cm]{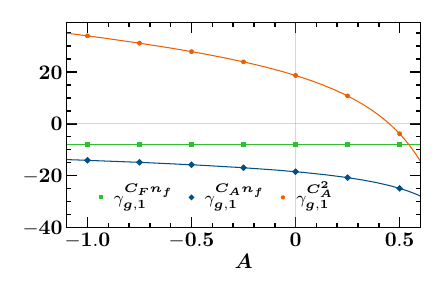}}
	\vspace{-2mm}
	\caption{\small{Two-loop non-cusp anomalous dimension $\gamma_{i,1}^B$ for the quark channels (left) and the gluon channels (right) as a function of the angularity $A$. The dots show the result of our numerical approach, and the lines are obtained by integrating the semi-analytical expressions from~\cite{Bell:2018vaa}. The various colour coefficients are defined in~\eqref{eq:non-cusp:colour}.}}
	\label{fig:non-cusp}
\end{figure}

As the N-jettiness beam functions are known analytically to the considered NNLO and beyond~\cite{Stewart:2010qs,Berger:2010xi,Gaunt:2014xga,Gaunt:2014cfa,Ebert:2020unb,Baranowski:2022vcn}, we can use these  expressions to check our results and to test the accuracy of our numerical routines. To do so, we first consider the two-loop non-cusp anomalous dimensions, which we decompose in terms of their colour structures according to
\begin{align} 
	\gamma_{q,1}^B &=
	\gamma_{q,1}^{C_F} \,C_F^2
	+ \gamma_{q,1}^{C_A} \,C_F C_A 
	+ \gamma_{q,1}^{n_f} \,C_F T_F n_f\,,
	\nonumber\\
	\gamma_{g,1}^B &= 
	\gamma_{g,1}^{C_A^2} \,C_A^2 
	+ \gamma_{g,1}^{C_A n_f} \,C_A T_F n_f
	+ \gamma_{g,1}^{C_F n_f} \,C_F T_F n_f\,.
	\label{eq:non-cusp:colour}
\end{align}
Specifically, we obtain for the jettiness ($A=0$) quark and gluon anomalous dimensions,
\begin{align}
	\gamma_{q,1}^{C_F}&= 21.2204 (228)\, & &\qquad [21.2203]\,,
	&\qquad \gamma_{g,1}^{C_A^2} &= 18.5940 (334)\, & &\qquad [18.5937]\,, 
	\nonumber	\\
	\gamma_{q,1}^{C_A}&= -6.5198 (342)\, & &\qquad [-6.5203]\,,
	&\qquad \gamma_{g,1}^{C_A n_f}  &= -18.4864 (66)\, & &\qquad [- 18.4863]\,, 
	\nonumber	\\
	\gamma_{q,1}^{n_f} &= -26.6988 (234)\, & &\qquad [- 26.6989]\,,
	&\qquad \gamma_{g,1}^{C_F n_f} &= -8.0000(1)\, & &\qquad [-8]\,, 
		\label{eq:non-cusp:jettiness}	
\end{align}
which are in excellent agreement with the known results that are shown in the square brackets. For other values of $A$, we can compare our numbers against the semi-analytical expressions for the soft anomalous dimension provided in~\cite{Bell:2018vaa} using RG consistency relations. Our results for seven values of the angularity $A \in \{-1, -0.75, -0.5, -0.25, 0, 0.25, 0.5\}$ are indicated by the dots for the quark channels (left panel) and the gluon channels (right panel) in Fig.~\ref{fig:non-cusp}, and they are again in perfect agreement with the existing results (solid lines). As can be read off from the numbers in \eqref{eq:non-cusp:jettiness}, the uncertainties of our numerical results are typically of a few per mil and they are therefore not visible on the scale of the plots. The agreement in Fig.~\ref{fig:non-cusp} provides a strong check of our novel automated SCET-1 setup.

%%%%%%%%%%%%%%%%%%%%%%%%%%%%%%%%%%%%%%%%%%%%%%%%%%%%%%%%%%%%%%%%%%%%%%%%
%%%%%%%%%%%%%%%%%%%%%%%%%%%%%%%%%%%%%%%%%%%%%%%%%%%%%%%%%%%%%%%%%%%%%%%%

We next turn to the renormalised matching kernels $I_{i\leftarrow j}^{(2)}(z)$ and address their distribution structure. Specifically, we find that the coefficients of the various plus distributions are fixed by known ingredients, and they therefore do not provide any new information. For a generic SCET-1 observable, we obtain the representation 
\begin{align}
&	I_{i\leftarrow j}^{(2)}(z) = 
	\delta_{ij} \, \bigg\{ 
	\frac{2 (\Gamma_0^i)^2}{g(n)^2} \left[\frac{\ln^3 (1-z)}{1-z}\right]_+ \!
	+ \frac{\Gamma_0^i (6 \gamma_{i,0}^\phi - 3 \gamma_{i,0}^B - 2 \beta_0)}{g(n)^2} \left[\frac{\ln^2 (1-z)}{1-z}\right]_+ \!
		\nonumber\\[0.2em]
	&\quad
	+ \bigg( \frac{2 \Gamma_1^i}{g(n)} + \frac{(2 \gamma_{i,0}^\phi - \gamma_{i,0}^B - 2 \beta_0) (2 \gamma_{i,0}^\phi - \gamma_{i,0}^B)}{g(n)^2} - 
	\frac{2\pi^2}{3} \frac{(\Gamma_0^i)^2}{g(n)^2} 
	+ \frac{2  \,\calC^{\delta}_{i,1} \Gamma_0^i}{g(n)} \bigg)  \left[\frac{\ln (1-z)}{1-z}\right]_+ \!
		\nonumber\\[0.2em]
	&\quad
	+ \bigg( 
	\frac{2 \gamma_{i,1}^\phi - \gamma_{i,1}^B}{g(n)} 
	- \frac{\pi^2}{3} \frac{\Gamma_0^i (2 \gamma_{i,0}^\phi - \gamma_{i,0}^B)}{g(n)^2} 
	+ \frac{4\zeta_3 (\Gamma_0^i)^2}{g(n)^2} 
	+ \frac{\calC^{\delta}_{i,1}(2 \gamma_{i,0}^\phi - \gamma_{i,0}^B - 2 \beta_0)}{g(n)} \bigg)  \left[\frac{1}{1-z}\right]_+ \!
		\nonumber\\[0.2em]
&\quad
	+ \calC^{\delta}_{i,2} \, \delta(1-z) \bigg\}
	+  {I}_{i\leftarrow j}^{(2,\grid)}(z)\,,
	\label{eq:I2ij:distributions}
\end{align}
which in addition to the anomalous dimensions that appear in the RG equation \eqref{eq:rgeI} depends on two novel quantities. These are the coefficients of the delta distribution in the NLO matching kernels \eqref{eq:I1ij}, which for the angularities is given by
\begin{align}
	\calC^{\delta}_{i,1} &= C_i \, \frac{A(4-A)}{(1-A)(2-A)}\, \frac{\pi^2}{6}\,,
\end{align}
and the corresponding coefficients of the delta distribution in the splitting functions $\gamma_{i,m}^\phi$. For convenience, we provide the latter coefficients in App.~\ref{app:anoD} as well.

%%%%%%%%%%%%%%%%%%%%%%%%%%%%%%%%%%%%%%%%%%%%%%%%%%%%%%%%%%%%%%%%%%%%%%%%
\begin{table}[t!]
	\center
	\setlength{\extrarowheight}{5pt}
	\scalebox{0.85}{\begin{tabular}{|c||c|c|c|c||c|c|c|}
			\hline
			$A$ & $-1$  &  $-0.75$ & $-0.5$  & $-0.25$ &  $0.25$  &  $0.5$ \\[6pt]
			\hline \hline
			$\calC_{q,2}^{C_F}$ & $-2.908(31)$ & $-2.764(36)$ & $-2.629(54)$ & $-2.494(243)$ & $-0.922(136)$ & $6.407(131)$ 
			\\[6pt]
			\hline
			$\calC_{q,2}^{C_A}$  &$-4.097(234)$ & $-7.189(231)$ & $-9.832(244)$&$-11.842(368)$ &$-11.207(232)$ & $-2.687(256)$
			\\[6pt]
			\hline
			$\calC_{q,2}^{n_f}$  &$6.320(24)$ &$5.153(33)$ &$3.690(35)$ &$1.717(37)$ &$-6.365(48)$ & $-17.488(62)$
			\\[6pt]
			\hline
	\end{tabular}}
	\caption{\small{NNLO coefficient of the delta distribution defined in \eqref{eq:I2ij:distributions} for the quark channel.}}
	\label{tab:Cdelta}
\end{table}
%%%%%%%%%%%%%%%%%%%%%%%%%%%%%%%%%%%%%%%%%%%%%%%%%%%%%%%%%%%%%%%%%%%%%%%%

The coefficient of the delta distribution in the last line of \eqref{eq:I2ij:distributions}, on the other hand, represents a genuine NNLO coefficient that we determine for the first time in this work (for $A\neq 0$). Using a similar colour decomposition as in \eqref{eq:non-cusp:colour}, 
\begin{align} 
	\calC^{\delta}_{q,2} &=
	\calC_{q,2}^{C_F} \,C_F^2
	+ \calC_{q,2}^{C_A} \,C_F C_A 
	+ \calC_{q,2}^{n_f} \,C_F T_F n_f\,,
	\nonumber\\
	\calC^{\delta}_{g,2} &= 
	\calC_{g,2}^{C_A^2} \,C_A^2 
	+ \calC_{g,2}^{C_A n_f} \,C_A T_F n_f
	+ \calC_{g,2}^{C_F n_f} \,C_F T_F n_f\,,
	\label{eq:delta:colour}
\end{align}
we can again compare our numbers for the specific jettiness observable ($A=0$) to the analytical results provided in~\cite{Gaunt:2014xga,Gaunt:2014cfa}. This yields a similar pattern,
\begin{align}
	\calC_{q,2}^{C_F} &=-2.165(47) \, & &\qquad [-2.165]\,,
	&\qquad 
	\calC_{g,2}^{C_A^2} &=-14.894(150)  \, & &\qquad [-14.897]\,, 
	\nonumber	\\
	\calC_{q,2}^{C_A}&=-12.732(125)  \, & &\qquad [-12.732]\,,
	&\qquad 
	\calC_{g,2}^{C_A n_f}  &=-1.237(33) \, & &\qquad [- 1.238]\,, 
	\nonumber	\\
	\calC_{q,2}^{n_f} &=-1.237(41) \, & &\qquad [-1.238]\,,
	&\qquad 
	\calC_{g,2}^{C_F n_f} &=0 \, & &\qquad [0]\,,
	\label{eq:delta:jettiness}	
\end{align}
showing once again excellent agreement between our numerical results and the available numbers shown in the brackets. The corresponding numbers for the remaining six angularities are new, and they are summarised in Tab.~\ref{tab:Cdelta}. As these coefficients obey Casimir scaling, it is in fact sufficient to provide the numbers for the quark channels only.

%%%%%%%%%%%%%%%%%%%%%%%%%%%%%%%%%%%%%%%%%%%%%%%%%%%%%%%%%%%%%%%%%%%%%%%%
\begin{figure}[t!]
	\centerline{
		\includegraphics[height=5.1cm]{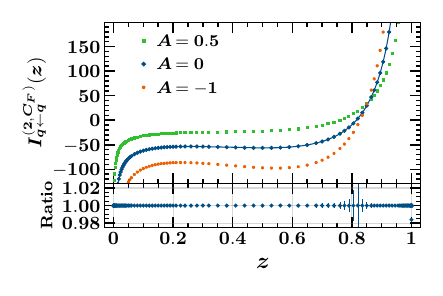}
		\includegraphics[height=5.1cm]{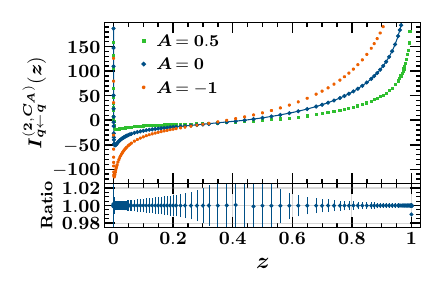}}
	\vspace{-2mm}
	\centerline{
		\includegraphics[height=5.1cm]{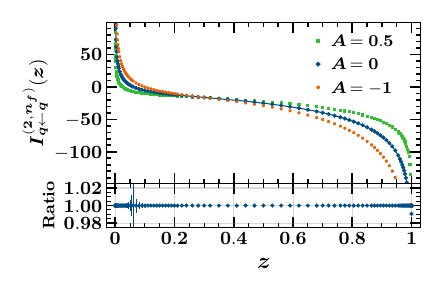}
		\includegraphics[height=5.1cm]{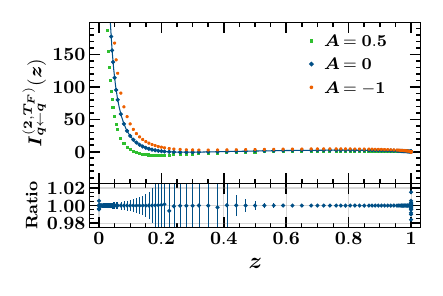}}
	\vspace{-2mm}
	\centerline{
		\includegraphics[height=5.1cm]{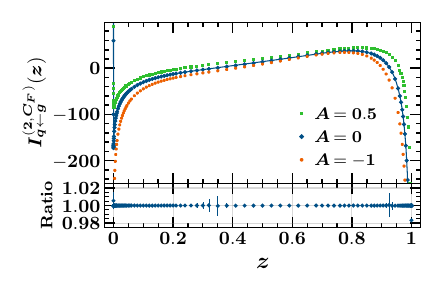}
		\includegraphics[height=5.1cm]{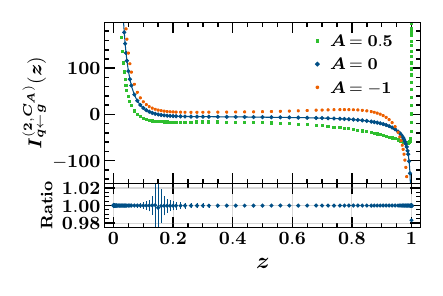}}
	\vspace{-2mm}
	\centerline{
		\includegraphics[height=5.1cm]{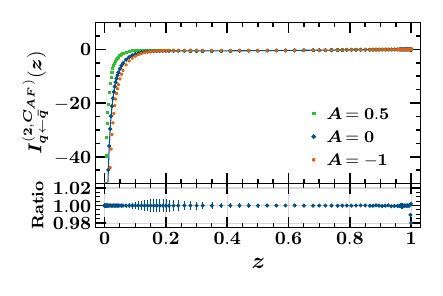}}
	\vspace{-2mm}
	\caption{\small{Grid contributions to the NNLO quark matching kernels  defined in \eqref{eq:NNLOkernels} for three values of the angularity $A$. The solid (blue) lines show the analytical jettiness results ($A=0$) from~\cite{Gaunt:2014xga,Gaunt:2014cfa}, and the lower panels display the ratio between our numbers and the analytical results. }}
	\label{fig:grids:quark}
\end{figure}
%%%%%%%%%%%%%%%%%%%%%%%%%%%%%%%%%%%%%%%%%%%%%%%%%%%%%%%%%%%%%%%%%%%%%%%%

%%%%%%%%%%%%%%%%%%%%%%%%%%%%%%%%%%%%%%%%%%%%%%%%%%%%%%%%%%%%%%%%%%%%%%%%
\begin{figure}[t!]
	\centerline{
		\includegraphics[height=5.1cm]{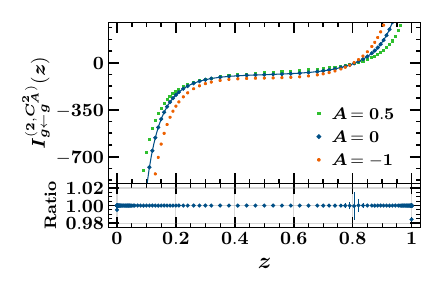}
		\includegraphics[height=5.1cm]{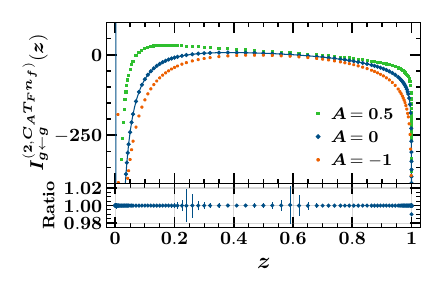}}
	\vspace{-2mm}
	\centerline{
		\includegraphics[height=5.1cm]{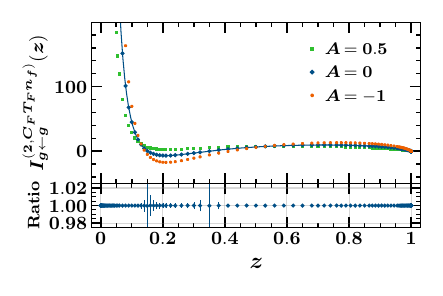}
		\includegraphics[height=5.1cm]{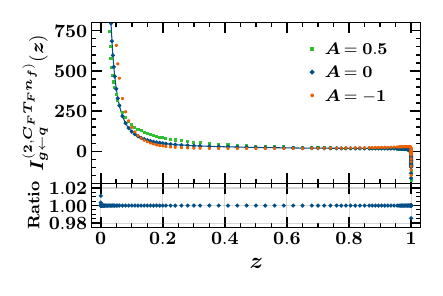}}
	\vspace{-2mm}
	\centerline{
		\includegraphics[height=5.1cm]{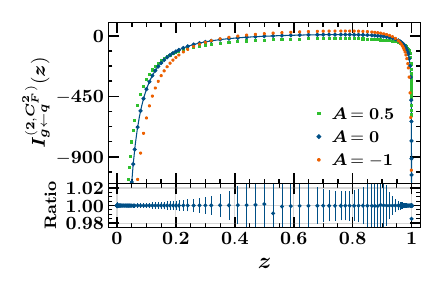}
		\includegraphics[height=5.1cm]{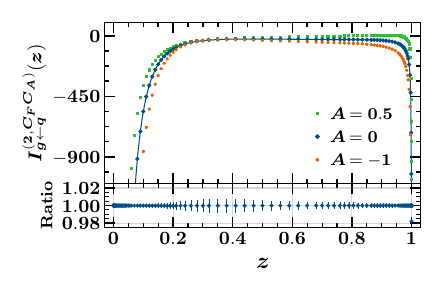}}
	\vspace{-2mm}
	\caption{\small{The same as in Fig.~\ref{fig:grids:quark} for the NNLO gluon kernels defined in \eqref{eq:NNLOkernels}.}}
	\label{fig:grids:gluon}
\end{figure}
%%%%%%%%%%%%%%%%%%%%%%%%%%%%%%%%%%%%%%%%%%%%%%%%%%%%%%%%%%%%%%%%%%%%%%%%

We finally turn to the (non-distributional) grid contributions defined in \eqref{eq:I2ij:distributions}, which we decompose similarly in terms of their colour coefficients,
\begin{align}
	{I}_{q\leftarrow q}^{(2,\grid)}(z)&=
	C_F^2 \; {I}_{q\leftarrow q}^{(2,C_F)}(z)
	+ C_F C_A \; {I}_{q\leftarrow q}^{(2,C_A)}(z)   
	+ C_F T_F n_f \; {I}_{q\leftarrow q}^{(2,n_f)}(z)
	+ C_F T_F \; {I}_{q\leftarrow q}^{(2,T_F)}(z)\,,
	\nonumber\\[0.2em]
	{I}_{q\leftarrow g}^{(2,\grid)}(z) &=
	C_F T_F \; {I}_{q\leftarrow g}^{(2,C_F)}(z) 
	+ C_A T_F \; {I}_{q\leftarrow g}^{(2,C_A)}(z) \,,
	\nonumber\\[0.2em]
	{I}_{q\leftarrow \bar q}^{(2,\grid)}(z) &=
	C_F (C_A - 2 C_F) \; {I}_{q\leftarrow \bar q}^{(2,C_{AF})}(z)  
	+ C_F T_F \; {I}_{q\leftarrow q}^{(2,T_F)}(z) \,,
	\nonumber\\[0.2em]
	{I}_{q\leftarrow q'}^{(2,\grid)}(z) &=
	{I}_{q\leftarrow \bar q'}^{(2,\grid)}(z) =
	C_F T_F \; {I}_{q\leftarrow q}^{(2,T_F)}(z)\,,
	\nonumber\\[0.2em]
	{I}_{g\leftarrow g}^{(2,\grid)}(z) &=
	C_A^2 \; {I}_{g\leftarrow g}^{(2,C_A^2)}(z) 
	+ C_A T_F n_f\; {I}_{g\leftarrow g}^{(2,C_A T_F n_f)}(z) 
	+ C_F T_F n_f \; {I}_{g\leftarrow g}^{(2,C_F T_F n_f)}(z) \,,
	\nonumber\\[0.2em]
	{I}_{g\leftarrow q}^{(2,\grid)}(z) &=
	C_F T_F n_f \; {I}_{g\leftarrow q}^{(2,C_F T_F n_f)}(z) 
	+ C_F^2 \; {I}_{g\leftarrow q}^{(2,C_F^2)}(z) 
	+ C_F C_A \; {I}_{g\leftarrow q}^{(2,C_F C_A)}(z) \,,
	\label{eq:NNLOkernels}
\end{align}
where $q'$ and $\bar q'$ indicate that the flavour of the incoming parton is different from the outgoing one. In total, there are thus seven independent matching kernels for the quark beam function, and another six kernels for the gluon beam function, which we sampled numerically for seven values of the angularity $A \in \{-1, -0.75, -0.5, -0.25, 0, 0.25, 0.5\}$ and 127 values of the momentum fraction $z$. The result is displayed for three of the angularities in Fig.~\ref{fig:grids:quark} (quark kernels) and Fig.~\ref{fig:grids:gluon} (gluon kernels). Our final numbers for all seven angularities, as well as their uncertainties that are not visible in the plots, are also provided in electronic form in the file that accompanies the present article.

%%%%%%%%%%%%%%%%%%%%%%%%%%%%%%%%%%%%%%%%%%%%%%%%%%%%%%%%%%%%%%%%%%%%%%%%
\begin{figure}[t!]
\centerline{
\hspace{-2mm}
\includegraphics{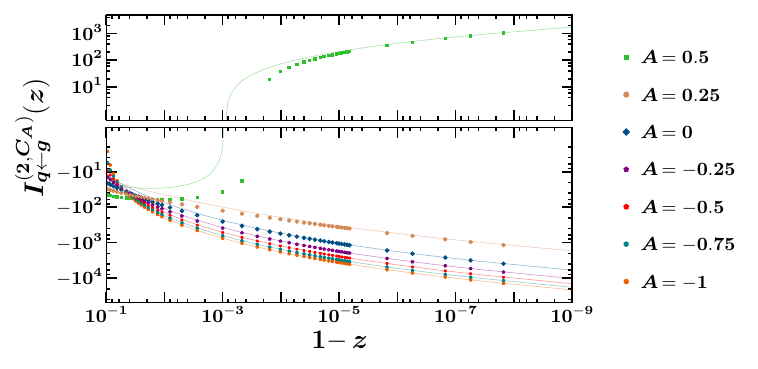}
}
%\vspace{-2mm}
\caption{\small{Close-up of the endpoint region $z\to 1$ for the ${I}_{q\leftarrow g}^{(2,C_A)}(z)$ grid contribution. Our numerical results are indicated by the dots and the solid lines refer to the asymptotic expression \eqref{eq:Iqg2CA:endpoint}, for which only logarithmic terms are included}.}
\label{fig:Iqg2CA:endpoint}
\end{figure}
%%%%%%%%%%%%%%%%%%%%%%%%%%%%%%%%%%%%%%%%%%%%%%%%%%%%%%%%%%%%%%%%%%%%%%%%

As a final check of our results, we show the analytical jettiness numbers (for $A=0$) as a solid (blue) line in each of the plots. The ratio between our numbers and the analytical results is furthermore displayed as a lower panel in each plot, which clearly shows that the accuracy of our numerical results is excellent, reaching a few percent only in those regions, in which the central values themselves are very small. Moreover, one observes that most of the kernels diverge at both endpoints $z\to 0$ and $z\to 1$, and the latter behaviour is particularly interesting for the off-diagonal ${I}_{q\leftarrow g}^{(2,C_A)}(z)$ kernel, since the curve asymptotes to $+\infty$ for $A=0.5$ in this case, whereas it points into the opposite direction for $A=0$ and $A=-1$. In order to verify if this is the correct scaling in the threshold limit, we performed an analytical study for this particular matching kernel, yielding
\begin{align}
\label{eq:Iqg2CA:endpoint}
	{I}_{q\leftarrow g}^{(2,C_A)}(z) \!&=\!
    \frac{ 2 (4 - 16 A + 13 A^2)}{3(2 - A)^2} \ln^3(1-z)
    \!+\! \bigg( \frac{2 (2 - 3 A)}{2 - A} + \frac{4 A (5 - 3 A)}{(2 - A)^2} \, \frac{\pi^2}{3} \bigg) \ln(1 - z) \!+\! \ldots
\end{align}
up to terms that are not logarithmically enhanced (or power-suppressed) in the threshold limit. We remark that these corrections represent a next-to-leading power effect, since the dominant terms in the threshold limit are captured by the distributions in \eqref{eq:I2ij:distributions}. Interestingly, we indeed find that the coefficient of the leading logarithm flips its sign around $A\simeq 0.35$. The endpoint region is furthermore displayed for this particular kernel in Fig.~\ref{fig:Iqg2CA:endpoint}, which shows that our numerical results are perfectly stable down to very low values of $(1-z)$. Further details on our analytical study of the threshold logarithms will be presented in a future publication~\cite{BBD}.

%%%%%%%%%%%%%%%%%%%%%%%%%%%%%%%%%%%%%%%%%%%
\section{Conclusion}
%%%%%%%%%%%%%%%%%%%%%%%%%%%%%%%%%%%%%%%%%%%
\label{sec:conclusion}

We computed the beam-function matching kernels for angularity distributions to NNLO in QCD. To this end, we transformed our automated setup that we previously developed for jet-veto observables~\cite{Bell:2022nrj,Bell:2024epn} to the generic SCET-1 case. In general, the non-logarithmic contributions to the matching kernels consist of distributions, see \eqref{eq:I2ij:distributions}, and a `grid' contribution that we sampled numerically for the full set of matching kernels relevant for both quark- and gluon-induced processes for seven values of the angularity and 127 values of the momentum fraction $z$. Our final results are displayed in Fig.~\ref{fig:grids:quark} for the quark kernels and in Fig.~\ref{fig:grids:gluon} for the gluon kernels, and they are also provided in electronic form as supplementary material that accompanies this paper.

We performed several checks to test the correctness of our numerical results. First, we verified that the divergence (and distribution) structure is correctly reproduced to high accuracy (typically on the sub-percent level). Moreover, we compared our numbers for one of the angularities ($A=0$) against the existing jettiness formulae~\cite{Gaunt:2014xga,Gaunt:2014cfa}, showing once again perfect agreement for all channels. Finally, we performed a novel study to extract the logarithmic rise for one exemplary kernel in the threshold limit $z\to 1$, which revealed that our numbers are stable even in the deep endpoint region. We believe that a dedicated analysis of the endpoint regions can provide useful supplementary information, and we plan to elaborate on this point in more detail in a future publication.

With the current work, our automated framework for sampling NNLO beam-function matching kernels directly in momentum space is now complete. For the specific case of DIS angularity distributions, our calculation provides the last ingredient to extend the resummation to NNLL$'$ accuracy. In view of the flexibility of our approach and the rising interest in more sophisticated DIS and hadron-collider observables, we anticipate many further applications of our framework in the future. In the long term, we plan to publish a code for the computation of NNLO beam-function matching kernels in the spirit of the \href{https://softserve.hepforge.org}{{\tt SoftSERVE}} distribution.

\acknowledgments
We thank D.\ Kang and J.\ Zhu for helpful discussions. G.D.\ thanks the \texttt{ZIMT} support for the \texttt{OMNI} cluster of the University of Siegen where the computation was performed. This work was supported by the Deutsche Forschungsgemeinschaft (DFG, German Research Foundation) under grant 396021762 - TRR 257 (\emph{``Particle Physics Phenomenology after the Higgs Discovery''}).

\begin{appendix}

%%%%%%%%%%%%%%%%%%%%%%%%%%%%%%%%%%%%%%%%%%%
\section{Anomalous dimensions and convolutions}
%%%%%%%%%%%%%%%%%%%%%%%%%%%%%%%%%%%%%%%%%%%
\label{app:anoD}

The lowest-order coefficients of the cusp anomalous dimension defined in \eqref{eq:AD} read
\begin{align} 
	\Gamma_0^i &= 4 C_i\,,
	\nonumber \\
	\Gamma_1^i &= 4 C_i
	\left\{ \left(\frac{67}{9}-\frac{\pi^2}{3}\right) C_A
	- \frac{20}{9} T_F n_f \right\}, 
\end{align}
where $i=F$ refers to the fundamental and $i=A$ to the adjoint representation. The non-cusp beam anomalous dimension, on the other hand, can be derived using the consistency relation $\gamma^{B}_i=-\frac12 (\gamma^{H}_i + \gamma^{S}_i)$ from the known values of the hard and soft anomalous dimensions. In the notation of \eqref{eq:AD}, the former are given by~\cite{Becher:2006mr}
\begin{align} 
	\gamma_{q,0}^H &= \!-12 C_F\,,
	\\
	\gamma_{q,1}^H &=\!
	C_F^2 \Big( \!- 6 + 8\pi^2 -96\zeta_3\Big)\!
	+ C_F C_A \bigg(\!-\frac{1922}{27} - \frac{22\pi^2}{3} + 104\zeta_3\bigg) \!
	+ C_F T_F n_f \bigg(\frac{520}{27}+\frac{8\pi^2}{3} \bigg)
	\nonumber
\end{align}
for the quark channel, and~\cite{Becher:2009qa}
\begin{align} 
	\gamma_{g,0}^H &= -\frac{44}{3}C_A+\frac{16}{3} T_F n_f\,,
	\nonumber\\
	\gamma_{g,1}^H &= 
	C_A^2 \bigg( -\frac{2768}{27} + \frac{22\pi^2}{9} +8\zeta_3\bigg)
	+ C_A T_F n_f \bigg(\frac{1024}{27}-\frac{8\pi^2}{9} \bigg) 
	+ 16\,C_F T_F n_f 
\end{align}
for the gluon channel. The non-cusp soft anomalous dimension has been determined to two-loop order in~\cite{Bell:2018vaa}. It reads
\begin{align} 
	\gamma_{i,0}^S &= 0\,,
	\\
	\gamma_{i,1}^S &= 	 \frac{2 C_i}{1-A} 
	\Bigg\{ \!\bigg(\!-\frac{808}{27} + \frac{11\pi^2}{9} + 28\zeta_3 - \Delta\gamma_1^{CA}(A)\bigg) C_A 
	+ \bigg(\frac{224}{27} - \frac{4\pi^2}{9}  - \Delta\gamma_1^{nf}(A)\bigg) T_F n_f \!\Bigg\}
		\nonumber
\end{align}
with two-dimensional integral representations
\begin{align}
	\Delta\gamma_1^{CA}(A) \!&=\! \int_0^1 \!\!\!\! dx\!\int_0^1 \!\!\!\! dy\,
	\frac{32 x^2(1\plus xy\plus y^2)\bigl[ x(1\plus y^2) + (x\plus y)(1\plus xy)\bigr]}{y(1-x^2)(x+y)^2(1+xy)^2}
	\ln\frac{(x^A+ xy) (x+ x^A y)}{x^A(1+ xy)(x+y)}, 
	\nonumber \\
	\Delta\gamma_1^{nf}(A) \!&=\!  \int_0^1 \!\! dx\!\int_0^1 \!\! dy\,
	\frac{64 x^2(1 + y^2)}{(1-x^2)(x+y)^2(1+xy)^2}
	\ln\frac{(x^A+ xy) (x+ x^A y)}{x^A(1+ xy)(x+y)},
\end{align}
that can easily be evaluated numerically for any value of the angularity $A<1$.\\

\noindent
The relevant coefficients of the splitting functions $P_{i\leftarrow j}^{(m)}(z)$ as well as explicit expressions for their convolutions \eqref{eq:def:convolution} can be found in our previous work~\cite{Bell:2024epn} and will not be repeated here. Instead, we list the (observable-specific) convolutions of the one-loop splitting functions with the one-loop matching kernels,
\begin{align}
	&   \Big( I_{q\leftarrow l}^{(1)} \otimes  P_{l\leftarrow q}^{(0)} \Big) (z) 
	\nonumber\\
	&\quad  = 
	\frac{A(4-A)}{(1-A)(2-A)}\, \frac{\pi^2}{6} \,C_F P_{q \leftarrow q}^{(0)}(z)
	\nonumber\\
	&\qquad 
	+C_F^2 \,\Bigg\{
	\frac{4(1-A)}{(2-A)} \bigg( 
	12 \left[\frac{\ln^2(1-z)}{1-z}\right]_+  \!
 	+ 6 \left[\frac{\ln(1-z)}{1-z}\right]_+  \!
 	- \frac{4 \pi^2}{3} \!\left[\frac{1}{1-z}\right]_+  \!
 	+ 8 \zeta_3 \,\delta(1 - z)
	\nonumber\\
	&\hspace{20mm}
	- (1 + z) \Big(2\Li_2(z) + 6 \ln^2(1 - z) - \pi^2 \Big)
	- \frac{8(1 + z^2)}{1 - z} \ln(1-z) \ln z
	\nonumber\\
	&\hspace{20mm}
	+ \frac{1 + 3 z^2}{1 - z} \ln^2 z 
	- \frac{1 + 6 z - z^2}{1 - z} \ln z  
	- (7 - z) \ln(1 - z) 
	+ 2 (1 - z) \bigg)
	\nonumber\\
	&\hspace{20mm}
	+ 2(1-z)\Big( 4\ln (1-z) 	-2 \ln z 	-1 	\Big) 
	\Bigg\}
	\nonumber\\
	&\quad\quad
	+ C_F T_F \Bigg\{
	\frac{8(1-A)}{3(2-A)} \bigg( 
	- (1 + z) \Big(6 \Li_2(z) +  3\ln^2 z - \pi^2 \Big) 
	+ (3 + 9 z + 4 z^2) \ln z
		\nonumber\\
	&\hspace{24mm}
	+ \frac{(1 - z) (4 + 7 z + 4 z^2)}{z} \ln(1 - z) 
	+ 1 - z \bigg)
	\nonumber\\
	&\hspace{24mm}
	-8 z \ln z + \frac{8 (1 - z) (1 - 2 z - 2 z^2)}{3 z} 	\Bigg\} ,
	\nonumber\\
	&   \Big( I_{q\leftarrow l}^{(1)} \otimes  P_{l\leftarrow g}^{(0)} \Big) (z) 
	\nonumber\\
	&\quad  = 
		\frac{A(4-A)}{(1-A)(2-A)}\, \frac{\pi^2}{6} \,C_F  P_{q \leftarrow g}^{(0)}(z)
		+ \beta_0 \, I_{q \leftarrow g}^{(1)}(z) 
		\nonumber\\
	&\qquad 	
	+ C_F T_F \Bigg\{
	\frac{4(1-A)}{(2-A)} \bigg( 
	\!-(1 - 2 z) \Big(2\Li_2(z) +  \ln^2 z - \frac{\pi^2}{3} \Big)
	- 2 \ln z + (1 - z) (3 - 7 z)
		\nonumber\\
	&\hspace{24mm}	
	+ 2 (1 - 2 z + 2 z^2) \ln^2 \Big( \frac{1-z}{z} \Big) 
	- 2 (1 - z) (2 - 3 z) \ln \Big( \frac{1-z}{z} \Big)  
	 \bigg)	
		\nonumber\\
	&\hspace{24mm}	
	-4 (1+2 z) \ln z - 4 (2-z-z^2)
	\Bigg\} 
	\nonumber\\
	&\qquad 	
	+ C_A T_F \Bigg\{
		\frac{8(1-A)}{3(2-A)} \bigg( \!
		- (1 + 4 z) \Big(6 \Li_2(z) +  3\ln^2 z  \Big) 
		+ (3 + 31 z^2) \ln z 
				- \frac{1 + 14 z - 15 z^2}{2}
		\nonumber\\
&\hspace{24mm}			
		+ 6 (1 - 2 z + 2 z^2) \ln(1-z) \ln \Big( \frac{1-z}{z} \Big)  
		+ \frac{(1 - z) (4 + 7 z + 31 z^2)}{z} \ln(	1 - z) 
		\nonumber\\
&\hspace{24mm}					
		+2 z (3 - z)   \pi^2 \!  \bigg) \!
		+ 16 z (1 - z) \ln (1-z) 
	- 32 z \ln z 
	+ \frac{8(1 - 3z- 15 z^2 + 17 z^3)}{3z}
	\Bigg\} ,
	\nonumber\\
	&   \Big( I_{g\leftarrow l}^{(1)} \otimes  P_{l\leftarrow q}^{(0)} \Big) (z) 
	\nonumber\\
	&\quad  = 
	\frac{A(4-A)}{(1-A)(2-A)}\, \frac{\pi^2}{6} \,C_A P_{g \leftarrow q}^{(0)}(z)
	+ 3 C_F \, I_{g \leftarrow q}^{(1)}(z) 
		\nonumber\\
	&\qquad 
	+ C_F^2 \Bigg\{
	\frac{4(1-A)}{(2-A)} \bigg( 
	\!-(2 - z)  \Big(2\Li_2(z) +  \ln^2 z \Big)
	+ \frac{4 (2 - 2 z + z^2)}{z} \ln(1-z) \ln \Big( \frac{1-z}{z} \Big)  
		\nonumber\\
	&\hspace{20mm}
	- \frac{2 (1 - z) (3 - 2 z)}{z} \ln(1 - z) 
	+ \frac{2 (3 - 4 z + 2 z^2)}{z} \ln z 
	- \frac{(4 - 6 z + 3 z^2)\pi^2}{3z}
		\nonumber\\
&\hspace{20mm}
	+ \frac{5 (1 - z)}{z} - 1 + z
	 \bigg)
	+ 8 z \ln (1-z)-4 z \ln z -4z +4
	\Bigg\}
	\nonumber\\
	&\qquad 
	+ C_A C_F \Bigg\{
	\frac{8(1-A)}{3(2-A)} \bigg( 
	(4 + z)  \Big(6\Li_2(z) - \pi^2 \Big)
	+  \frac{3(2 - 2 z + z^2)}{z} \ln^2 \Big( \frac{1-z}{z} \Big) 
			\nonumber\\
	&\hspace{24mm}
	+  3 (4 + z) \ln^2 z 
	- \frac{(1 - z) (31 + 7 z + 4 z^2)}{z} \ln \Big( \frac{1-z}{z} \Big)  
			\nonumber\\
&\hspace{24mm}
	- \frac{2 (11 + 3 z + 3 z^2)}{z} \ln z 
	+ \frac{(1 - z) (21 - 5 z)}{2z} \bigg)
	\Bigg\},
	\nonumber\\
	&   \Big( I_{g\leftarrow l}^{(1)} \otimes  P_{l\leftarrow g}^{(0)} \Big) (z) 
	\nonumber\\
	&\quad  =
	\frac{A(4-A)}{(1-A)(2-A)}\, \frac{\pi^2}{6} \,C_A P_{g \leftarrow g}^{(0)}(z)
	\nonumber\\
	&\qquad 
	+\beta_0 C_A \,\Bigg\{
	\frac{8(1-A)}{(2-A)}  \bigg( 
	\left[\frac{\ln(1-z)}{1-z}\right]_+   \!
	-  \frac{z^3-(1-z)^2}{z} \,\ln \Big( \frac{1-z}{z} \Big) 
	- \frac{1}{1-z} \,\ln z \bigg) \Bigg\}
	\nonumber\\
	&\qquad
	+C_A^2 \,\Bigg\{
	\frac{4(1-A)}{(2-A)} \bigg( 
	12 \left[\frac{\ln^2(1-z)}{1-z}\right]_+  \!
	- \frac{4 \pi^2}{3} \!\left[\frac{1}{1-z}\right]_+  \!
	+ 8 \zeta_3 \,\delta(1 - z)
	+16 (1 + z) \,\Li_2(z)
	\nonumber\\
	&\hspace{18mm}
	+ \frac{12 (1 - 2 z + z^2 - z^3)}{z} \ln^2 (1-z)
	- \frac{8 (1 - z) (11 + 2 z + 11 z^2)}{3z} \ln(1-z) 
	\nonumber\\
	&\hspace{18mm}
	- \frac{16 (1 - z + z^2)^2}{z(1 - z)} \ln(1-z)\ln z 
	+ \frac{4 (11 - 21 z + 6 z^2 - 22 z^3)}{3z} \ln z 
	\nonumber\\
	&\hspace{18mm}
	+ \!\frac{4 (1 + 3 z^2 - 4 z^3 + z^4)}{z(1 - z)} \ln^2 z 
	- \!\frac{4 (1 + 3 z^2 - z^3)\pi^2}{3z} 
	+ \!\frac{2 (1 - z) (67 - 2 z + 67 z^2)}{9z} \! \bigg)\!
	\Bigg\}
	\nonumber\\
	&\quad\quad
	+ C_F T_F n_f \Bigg\{
	\frac{8(1-A)}{3(2-A)} \bigg( 
	- (1 + z) \Big(6 \Li_2(z) +  3\ln^2 z - \pi^2 \Big) 
		+ 3 \ln z 
	\nonumber\\
	&\hspace{28mm}
	+ \frac{(1 - z) (4 + 7 z + 4 z^2)}{z} \ln \Big( \frac{1-z}{z} \Big) 
	- \frac{(1 - z) (13 + 10 z + 13 z^2)}{3z}
	\bigg)
	\nonumber\\
	&\hspace{28mm}
	+ 8 z \ln z +4+4z-8 z^2 	\Bigg\} ,
\end{align}
where the factor $n_f$ in the last term arises from a sum over massless quark flavours.\\

\noindent 
As can be read off from \eqref{eq:I2ij:distributions}, the distribution structure of the diagonal matching kernels is determined by the anomalous dimensions \eqref{eq:AD}, the NLO coefficient $\calC^{\delta}_{i,1}$ and the coefficients of the delta distribution in the splitting functions $\gamma_{i,m}^\phi$. Specifically, the latter arises in the decomposition of the splitting functions as
\begin{align}
	P_{i\leftarrow i}(z,\alpha_s) 
	=
	\Gamma_{\rm cusp}^{R_i}(\alpha_s) \left[\frac{1}{1-z}\right]_+ 
	+ 
	\gamma_i^{\phi}(\alpha_s)  \;\delta(1-z) + \dots
\end{align}
Expanding the anomalous dimension $\gamma_i^{\phi}(\alpha_s)$ in analogy to \eqref{eq:AD}, the required coefficients can be extracted from the literature on threshold resummation, e.g.~\cite{Becher:2007ty,Ahrens:2009cxz}, yielding
\begin{align} 
	\gamma_{q,0}^\phi &= 3 C_F\,,
	\nonumber\\
	\gamma_{q,1}^\phi &=
	C_F^2 \bigg(  \frac32 -2\pi^2 +24\zeta_3\bigg)\!
	+ C_F C_A \bigg(\frac{17}{6} + \frac{22\pi^2}{9} - 12\zeta_3\bigg) \!
	- C_F T_F n_f \bigg(\frac23+\frac{8\pi^2}{9} \bigg)\,,
	\nonumber\\
	\gamma_{g,0}^\phi &= \frac{11}{3}C_A-\frac{4}{3} T_F n_f\,,
	\nonumber\\
	\gamma_{g,1}^\phi &= 
	C_A^2 \bigg( \frac{32}{3} +12\zeta_3\bigg)
	- \frac{16}{3}\, C_A T_F n_f 
	-4\,C_F T_F n_f \,.
\end{align}

\end{appendix}

\bibliography{DISangularities}

\end{document}